\documentclass[twocolumn,superscriptaddress,amsmath,amssymb,aps,pra,nofootinbib]{revtex4}
\usepackage{amsmath}
\usepackage{amssymb}
\usepackage{graphicx}
\usepackage{times}
\large

\newcommand{\be}{\begin{equation}}
\newcommand{\ee}{\end{equation}}

\newcommand{\barl}{\begin{eqnarray}}
\newcommand{\ear}{\end{eqnarray}}
\newcommand{\non}{\nonumber}
\newcommand{\sech}{\mathrm{sech}}
\begin{document}
\title{Effects of spatial dispersion on Self--induced transparency in two--level media}
\author{Zoran Ivi\'c}
\affiliation{University of Belgrade, Vin\v ca Institute, PO Box 522, 11001 Belgrade, Serbia}
\affiliation{National University of Science and Technology MISiS, Leninsky prosp. 4, Moscow 119049, Russia}
\author{Dalibor \v Cevizovi\'c}
\author{\v Zeljko Pr\v zulj}
\affiliation{University of Belgrade, Vin\v ca Institute, PO Box 522, 11001 Belgrade, Serbia}
\author{Nikos Lazaridess}
\affiliation{National University of Science and Technology MISiS, Leninsky prosp. 4, Moscow 119049, Russia}
\affiliation{Crete Center for Quantum Complexity and Nanotechnology, Department of Physics, University of Crete, P. O. Box 2208, Heraklion 71003, Greece}
\author{G.P.Tsironis}
\affiliation{National University of Science and Technology MISiS, Leninsky prosp. 4, Moscow 119049, Russia}
\affiliation{Crete Center for Quantum Complexity and Nanotechnology, Department of Physics, University of Crete, P. O. Box 2208, Heraklion 71003, Greece}
\begin{abstract}
We study the effects of  dispersion in carrier waves on the properties of soliton  self--induced transparency (SIT)  in two level media. We found substantial impact of dispersion effects on typical SIT soliton features. For example, the degree of SIT pulse velocity slowing down (acceleration) is determined by the ratio of the incoming pulse frequency over atomic transition frequency - $x=\omega/\omega_0$. Specifically, an immediate pulse stopping is predicted for absorbing media when pulse duration time exceeds some critical value. In the sharp line limit stopping may emerge only for frequency ratio above unity, while for the inhomogeneously broadened systems it appears irrespective of the value of  $x$. Analysis performed on the basis of Mcall\& Hahn \textit{Area theorem} implies that pulse stopping is achieved when Ber's absorption coefficient approaches infinity, that is, pulse energy is fully absorbed in the medium. In the case of amplifying media super-luminal motion is predicted as in the case of resonance. However, there is a lowest value in the frequency ratio below which the pulse velocity tends to the sub-luminal region. These new features of the SIT phenomenon open novel ways on how it may be exploited for the control of  electromagnetic wave radiation in two-level media. This may be achieved by varying frequency ratio. 
\end{abstract}
\maketitle
\section{Introduction}
Propagation of short, intense, electromagnetic (EM) pulses, resonantly interacting with two--level atomic media, lead to the emergence of a number of remarkable \textit{coherent cooperative quantum} phenomena such as Dicke superradiance, self--induced transparency (SIT), electromagnetically induced transparency (EIT), photon--echo, coherent population oscillation ... \cite{sr1,sr2,aleb,sit1,sit2,sit3,nlphys,eit1,eit2,pheh3,titt,cpo1,cpo2}.
During the last two decades these phenomena attracted particular attention due to the potential for practical applications, i.e., 
for  realization of \textit{quantum memories} \cite{QM1,QM2} that are devices fundamental for the future technologies for quantum communication and processing \cite{comm1,comm2,qc4,devo,geo,para}. In that context the achievement of  control over the propagation of EM radiation by matter \cite{QMeng,slow1,zag1,slowrev}, and similarly, the manipulation of atoms, natural and the artificial ones (superconducting or quantum dot qubits), by light  \cite{qbmanip},   became a really important issue  in physics. 

Great practical successes were achieved using EIT which turns out to be very useful allowing substantial slowing down  and even stopping of \textit{light pulses} in media composed of natural atoms (atomic vapors \cite{frez2,comm2,frez3,frez4,frez5,frez6,frez7,frezth1,frezth2}). In this way, the information carried by the pulse may be temporarily transferred to the medium. Pulses can then be 'revived' with their original information intact. Nevertheless, EIT -- based techniques have certain limitations in potential practical applications due to the narrow transparency spectrum \cite{OEX1} and vulnerability due to inevitable coupling with the environment leading to relaxation processes (\textit{homogeneous broadening}) and dephasing (\textit{inhomogeneous broadening}). Relaxation effects may be suppressed by exploiting high-intensity pulses. However, this may cause damage to the medium. For that reason, the development of  analogous techniques but adjusted for the microwave domain may be useful. These novel trends rely on engineered media, quantum metamaterials (QMM), built with artificial "atoms" made typically of superconducting 
circuits. This had motivated the investigations of a possible emergence of \textit{collective coherent quantum} phenomena in QMMs and their implementation in design of quantum technological devices  \cite{qbeit1,qbeit2,qbeit3,alu,arxiv,fqb,mand,scir}. The use QMMs had shown that, in parallel with EIT--based techniques, the use of SIT could be very useful. Specifically, employing SIT could make possible to phase out  out \textit{inhomogeneous broadening}, the main obstacle to quantum coherence, 
and actually turn it into an advantage. This is due to the fact inhomogeneous broadening is required for the emergence of SIT. We note that the whole concept of SIT relies on inhomogeneous broadening since the  Area theorem that is fundamental relies on it. Additionally,  relaxation effects (homogeneous broadening) may be avoided using (ultra)short light pulses with the duration far less than both transverse and longitudinal relaxation times.  

To recall, SIT is a lossless propagation of an optical pulse through an otherwise opaque optical medium composed of a large number of \textit{inhomogeneously broadened} two-level atoms. According to the McCall -- Hahn \textit{area theorem}, which is the main theoretical result of the whole concept, pulses travel through the media with no gain or loss when their area $\theta(x)\sim \int dt E(x,t) =n\pi$, where  $E(x,t)$ is the magnitude of electric field. For even (odd) $n$ pulses are stable (unstable). When 
the pulse  area is below $\pi$, the pulse gradually weakens with traveled distance and finally disappears being absorbed by the medium; on the other side, pulses whose input areas are slightly above $\pi$ increase their area up to a $2\pi$ after which point continue stable propagation as  $2\pi$--soliton. Finally, all pulses with the areas $n\pi$ for $n>2$ split in two, three, etc  solitons.  

While the area theorem is a purely theoretical result, obtained under very restrictive assumptions,  it
is nevertheless  surprising how accurately it describes most of the main features of SIT as confirmed by numerous experimental results \cite{aleb}. For that reason, at least concerning the original problem -- SIT in a system of \textit{atomic vapors}, it's further theoretical consideration could be superfluous. However, recent proposals \cite{alu,arxiv,fqb,mand,scir} of the practical applications of SIT effect in media built of artificial atoms, quantum metamaterials (QMM) for example, reopens the importance of theoretical studies especially those effects which have not been considered.  One such issue is the study of the dispersion effects, i.e. dependence of the carrier wave frequency on its wave vector -- $\omega = \omega (k)$. In the original context \cite{sit1,sit2,sit3,nlphys}  it was assumed that the dispersion law satisfies the simple relation $ \omega (k)= ck$, where $c$ is the speed of light in the medium. However, as shown recently in the context of SIT in QMM \cite{fqb,mand,scir} and exciton SIT \cite{tal1,tal2} this issue cannot be neglected easily and, under some circumstances, turns out to be of particular interest resulting in some very peculiar features.

In this article, we study the effects of dispersion on SIT within the original model introduced by McCall and Hahn \cite{sit1,sit2}. We base our study on a system of \textit{reduced Maxwell--Bloch equations} (RMBE) obtained from the original one by eliminating fast oscillating terms in accordance with \textit{slowly varying envelope and phase approximation} (SVEA) \cite{aleb} leading, in final instance, to well known solutions 
\cite{aleb,sit1,sit2,sit3,nlphys}. A brief and very instructive  pedagogical overview may be found in \cite{nlphys}. 
\section{Evaluation of the dispersion law}
The starting point of our analysis is RMBEs found through consistent application of SVEA. 

\barl\label{SE02A}
\begin{split}
	\dot{S}_x&=&-(\Delta+\dot{\phi})S_y,\hspace{8em} & \text{a)}\\
	\dot{S}_y&=&(\Delta+\dot{\phi})S_x+\frac{\kappa}{2}\mathcal{E} S_z,\hspace{4em} & \text{b)}\\
	\dot{S}_z &=&-\frac{\kappa}{2}\mathcal{E} S_y.\hspace{9em} & \text{c)}
\end{split}
\ear
Here, $S_i(x,t)$, $\mathcal{E}(x,t)$ and $\phi(x,t)$ are new \textit{slow} dynamical variables corresponding, respectively, to transformed atomic functions, envelope and phase of the EM pulse. More precisely, $S_x$ and $S_y$ correspond to 'dispersive' and 'absorptive' components of induced polarization of the medium:
$P(x,t)=\mathcal{N} d \langle s_x(z,t)\rangle$, where $ s_x$ -- quantum mechanical expectation value of a $x$ component of Pauli spin matrix ($\sigma_x$) in a state being the superposition of ground and the excited ones. In terms of 'slow' variables it reads $s_x=S_x\cos\Psi(x,t)+S_y\sin\Psi(x,t); \;\; \Psi(x,t)=kx-\omega t+\phi(x,t)$ ; $\mathcal{N}$ stays for the concentration of atoms, $d$ is a dipole transition matrix element, $\Delta=\omega_0-\omega(k)$ with $\omega$ and $k$ are the frequency and the wave vector of a carrier wave and $\omega_0$ corresponding to atom transition frequency. The angular brackets in the last system refer to the fact that in practice we deal with a system of the inhomogeneously broadened atoms. That is, $\omega_0$ corresponds to an atomic  transition frequency that is \textit{different for each atom}. In the case of a system composed of a large number of atoms, all of these frequencies may be taken continuously distributed around some mean value. Since the large wavelength ($\lambda \gg d$) EM pulse interacts simultaneously with large number of IH broadened "atoms" the collective back--action on the  propagating pulse must be described in terms of the average "polarization" as follows--$\langle ....\rangle=\int_{0}^{\Delta}(....) \mathcal{G}(\Delta') d\Delta'$ where $\mathcal{G}(\Delta)$ is normalized to unity ($\int^{\infty}_0 d\Delta \mathcal{G}(\Delta)=1$) line--shape function. Finally,   
$c=c_0/n$ is the speed of light in TL medium with index of refraction $n$. In addition to the above system, there are two more equations  
arising as a result of the transformation of a single \textit{second order} equation to a two first order ones for the amplitude and phase. 

\begin{equation}\label{SE02B}
\begin{split}
\frac{c^2}{2\omega}\Big(k^2-\frac{\omega^2}{c^2}\Big)\mathcal{E}+\Big(\dot{\phi}+\frac{kc^2}{\omega}{\phi}'\Big)\mathcal{E}=\frac{\gamma\omega^2_0}{2\omega}\langle S_x\rangle,\qquad & \text{a)}\\
\dot{\mathcal{E}}+\frac{kc^2}{\omega}{\mathcal{E}}'=\frac{\gamma\omega^2_0}{2\omega}\langle S_y\rangle, \; \gamma = 4\pi \mathcal{N}d.\hspace{3em} & \text{b)}
\end{split}
\end{equation}

The necessity for the consideration of dispersion effects may be viewed on the basis of equations (2) and (3). We first recall that the simplest analytic  solutions of these equations exist in resonance $\omega = \omega_0$ and for stationary phase $\dot \phi\equiv \phi'=0.$ In that case the  system of equations (2) and (3) greatly simplifies and may be solved using trigonometric parameterization in terms of Bloch angle ($S_y=S_0\sin\theta$ and $S_z=S_0\cos\theta$) \cite{aleb,sit1,sit2,sit3,nlphys}, which, satisfies sine--Gordon equation. In addition, employing the usual initial conditions that population inversion  $S_0=S_z(-\infty)\equiv\pm 1$, from system (2), we  obtain that $S_x=const\equiv 0.$ Its immediate consequence is the dispersion law $\omega = \pm kc$ (see eq. 3.a) that  holds only at resonance. Out of resonance, this relation is only approximate and a correct treatment requires determination it's true form. In this case parametrization in terms of Bloch angles is again possible but demands an \textit{ad hock} assumption known as \textit{factorization ansatz} introducing  the \textit{spectral response function}: $S_y(\Delta)=F(\Delta)S_y(\Delta=0)$. This approach leads once again to the SG equation for the Bloch angle. Knowing that $)S_y(\Delta=0)=S_0\sin\theta$ and that $\dot{\theta}=-\frac{\kappa}{2}\mathcal{E}$ SG equation for Bloch--angle became
\be\label{SG} \ddot{\theta}+\frac{kc^2}{\omega}\dot{\theta'}=\frac{\gamma\omega^2_0\kappa}{4\omega}{\langle F(\Delta\rangle}\sin\theta. \ee
Solutions of this equation are well known, they are 
$2\pi$ solitons. For the spectral function we use known relation \cite{aleb,sit1,sit2} connecting it with the detuning $\Delta$ and pulse duration time $\tau_p$

However, eq. (4) and thus its solutions still contain a single yet undetermined parameter $k$ that  appears in the evaluation of the soliton delay ratio ($v/c$), absorption coefficient ($\alpha$) and the area theorem. In other words, all these functions are fucntions of $k$ whose explicit knowledge is required to examine the potential usability of SIT in practical applications. Thus, the relation (\ref{SG}) is useless in that respect. in  order find  the dispersion law, we need to go back to the system and assume that the pulse propagates undistorted in a soliton form. This enables us to take that all system variables depend on spatial coordinates and the time only through the variable $\tau=t-\frac{x}{v}$ (i.e. passing to moving frame). Accordingly, systems (\ref{SE02A}) and (\ref{SE02B}) become
\barl\label{mf1}
\begin{split}
	{S}_{x,\tau}&=&-(\Delta+\phi_{\tau})S_y,\hspace{8em} & \text{a)}\\
	{S}_{y,\tau}&=&(\Delta+{\phi}_{\tau})S_x+\frac{\kappa}{2}\mathcal{E} S_z,\hspace{4em} & \text{b)}\\
	S_{z,\tau} &=&-\frac{\kappa}{2}\mathcal{E} S_y.\hspace{9em} & \text{c)}
\end{split}
\ear
\begin{equation}\label{mf2}
\begin{split}
\Big({\phi}_{\tau}+\frac{G}{2\omega\Gamma}\Big)\mathcal{E}=\frac{\gamma\omega^2_0}{2\omega\Gamma}\langle S_x\rangle,\qquad & \text{a)}\\
\mathcal{E}_{\tau}=\frac{\gamma\omega^2_0}{2\omega\Gamma}\langle S_y\rangle, \hspace{5em} & \text{b)}\\
\Gamma=1-\frac{c^2 k}{\omega v}, \;\; G=\frac{\omega^2-c^2 k^2}{2\omega}. \;\;\;\; & \text{c)}
\end{split}
\end{equation}

The last two equations, combined with the third and first one of system \ref{mf1}, may be easily integrated to give $S_z$ and phase.  
It is trivial in the \textit{sharp line} limit, while for the finite broadening we have to employ the \textit{factorization ansatz} which enables one to expres $\langle S_y\rangle$ through $S_y$. For that purpose we took $S_y$ in factorized form and its average as: $\langle S_y(\Delta)\rangle =\langle F(\Delta)S_y(\Delta=0)\rangle \equiv \langle F(\Delta)\rangle S_y(\Delta = 0)\;\; \mathrm{with}\;\; F(\Delta)=\frac{1}{1+\Delta^2\tau^2_p}$.  Now we multiply  the last expression with $1=\frac{F(\Delta)}{F(\Delta)}$. This simple manipulation yields $\langle S_y\rangle = \frac{\langle F(\Delta)\rangle }{F(\Delta)}S_y$.

Employing this approach  in the last equation we may combine it with the third one in \ref{mf1} which finally yields:
\begin{equation}\label{ESz}
S_z=S_0-\frac{\omega\kappa\Gamma}{2\omega^2_0\gamma}\frac{F(\Delta)}{\langle F(\Delta)\rangle}\mathcal{E}^2.
\end{equation}

We note that $S_0$ is the  initial population of TLS where  $S_0=-1$ means that all TLS's is in their ground state, while $S_0=+1$ means that we have all TLS's  in the excited state.   

We may now  focus on the equation for phase--the first one in \ref{mf2}. We first differentiate it with respect to $\tau$, then we use first equation in \ref{mf1} to eliminate $\langle \dot S_x \rangle$. Also, in a final step we use same trik as above to express $\langle S_y\rangle$ through $S_y$. This finally yields :

\begin{equation}\label{EQ04}
\phi_{\tau\tau}\mathcal{E}+2\phi_{\tau}\mathcal{E}_{\tau}+
\Big(\tilde{\Delta}-\frac{G}{\Gamma}\Big)\mathcal{E}_{\tau}=0,\;\; \tilde \Delta=\frac{\langle\Delta F(\Delta)\rangle}{\langle F(\Delta) \rangle} \nonumber
\end{equation}

Its integration yields:

\begin{equation}\label{EQ06}
\phi_{\tau}=\frac{1}{2}\Big(\tilde{\Delta}-\frac{G}{\Gamma}\Big) \nonumber
\end{equation}

\noindent Here, we used the initial condition $\lim_{\tau\rightarrow -\infty}\mathcal{E}=0$. At this place, it is necessary to recall that the phase $\phi$ was introduced through overall phase $\psi(x,t)=kx-\omega t+\phi(x,t)$ in which linear terms in $x$ and $t$ are already accounted for independently of $\phi$, which, therefore, cannot contain terms \textit{linear} in ($x$ and $t$). That is, we must have take $\phi_{\tau}=0$. In such a way, last relation implies:

\begin{equation}\label{DR}
\tilde{\Delta}-\frac{G}{\Gamma}=0
\end{equation}  

This is quadratic equation for the pulse wave vector $k$. It may be solved for $k$ as function of $\tilde{\Delta}$, ratio $\omega/\omega_0$ and pulse velocity as a parameter. However, in a view of the analysis of experimental data, it is more convenient to examine $k$ in dependence of the pulse duration instead of its velocity. 

For that purpose we use known solutions: 
\barl\label{soliton}
\non\mathcal{E}&=&\mathcal{E}_0\sech(\tau/\tau_p)\\
\mathcal{E}_0&=&\sqrt{\frac{4\gamma S_0}{\kappa\omega\Gamma}\langle F(\Delta)\rangle},\; \;\mathcal{E}_0\tau_p=\frac{4}{\kappa},\ear
to obtain $\Gamma=\frac{\gamma \kappa S_0}{4\omega_0 x}\langle F(\Delta)\rangle \omega^2_0\tau_p^2 $ and eliminate it from (\ref{DR}). 
Our final results, expressed in terms of a dimensionless variables: $K=\frac{kc}{\omega_0}$-- the carrier wave quasimomentum,  the pulse velocity $V=\frac{v}{c}$ and frequency ratio $x=\frac{\omega}{\omega_0}$, read:
\barl\label{DL1}
\non K_{\pm}&=&\sqrt{x^2-2\frac{\tilde\Delta\nu S_0}{\omega_0}\langle F(\Delta)\rangle\omega^2_0\tau^2_p},\\
V_{\pm}&=&\frac{K}{ x-\nu S_0\langle F(\Delta)\rangle\omega^2_0\tau^2_p}.\ear
In the last expressions sign $+$ ($-$) corresponds to initial population inversion $S_0=-1$ ($S_0=1$). Here $\nu = \frac{\gamma\kappa}{4\omega_0}$, hereafter called material parameter, characterizes the strength of field--atom interaction. Apparently, the pulse propagation is determined both on its characteristics (duration time) and by the properties of the material. 

At this stage, before detailed discussion of the pulse propagation, we perform some preliminary calculations, in order to see how the initial conditions are reflected on the nature of the solution. To this end we derive an alternate form of  the dispersion law. 
\be\label{dlaw1}
K_{\pm}=-\frac{\tilde{\Delta}}{\omega_0 V}\pm \sqrt{(x+\frac{\tilde{\Delta}}{\omega_0})^2+\frac{\tilde{\Delta}^2}{\omega^2_0}(\frac{1}{V^2}-1)}.
\ee
From the expression for the pulse amplitude we immediately find that the existence of solutions requires $\Gamma S_0>0$. In the case of simple dispersion $\omega = ck$ it may be addressed to sub--luminal or super--luminal propagation for $S_0=-1$ or $S=1$. The same conclusion holds here and may be easily proved on the basis of (\ref{dlaw1}). Note that condition $\Gamma S_0>0$ may be rewritten as 
\be\label{cond} \Big(1- \frac{K}{xV}\Big)S_0>0.\ee
That is, $S_0 = -1$ requires $K>xV$, which, together (\ref{dlaw1}), after some straightforward calculation, yields $x^2(1-V^2)>0$ implying sub--luminal propagation. The same reasoning results with  $x^2(1-V^2)<0$ for $S_0=1$ leading to super--luminal motion.
\begin{figure}[h]
	\begin{center}
		\includegraphics[height=6 cm]{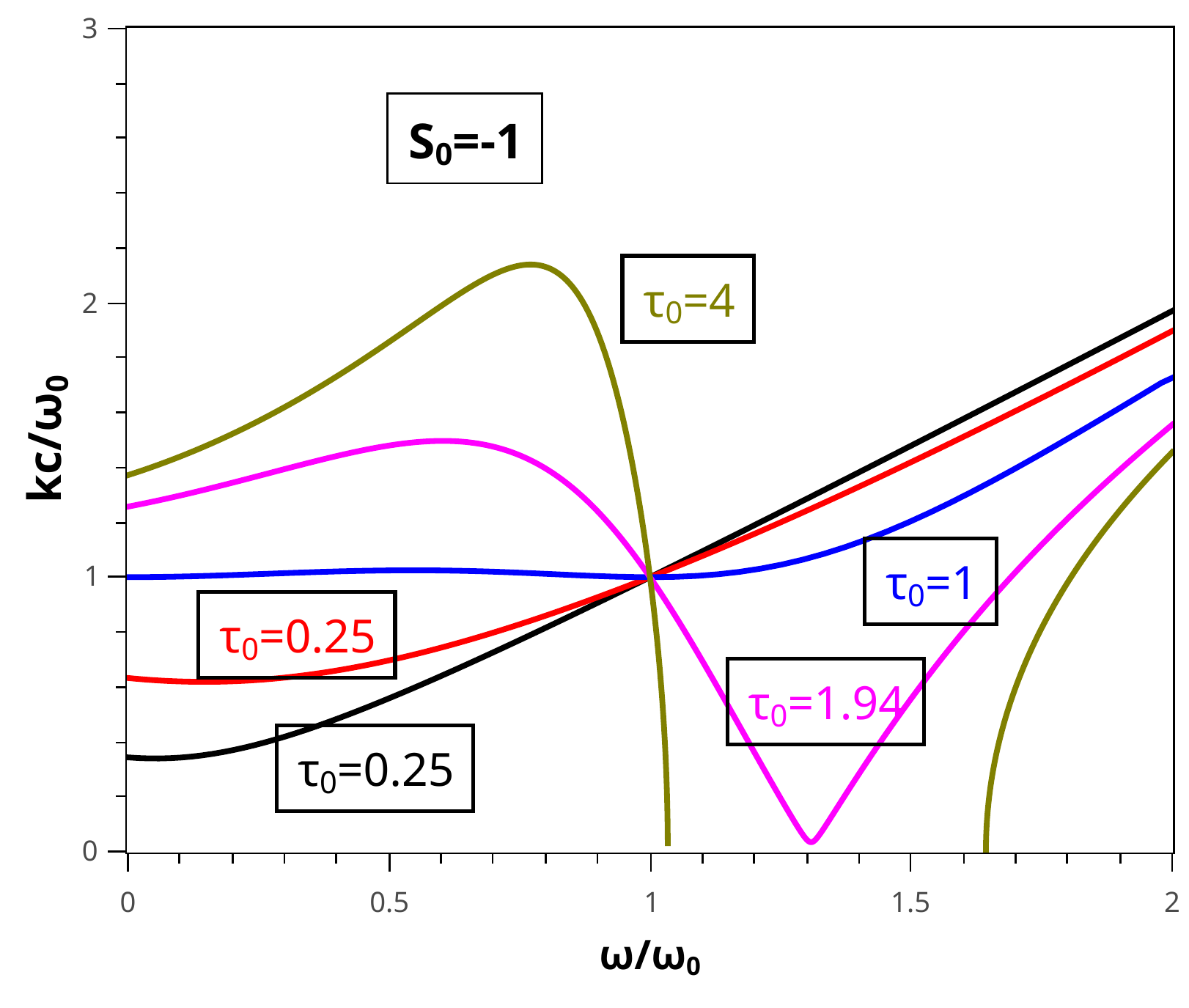}
		\includegraphics[height=6 cm]{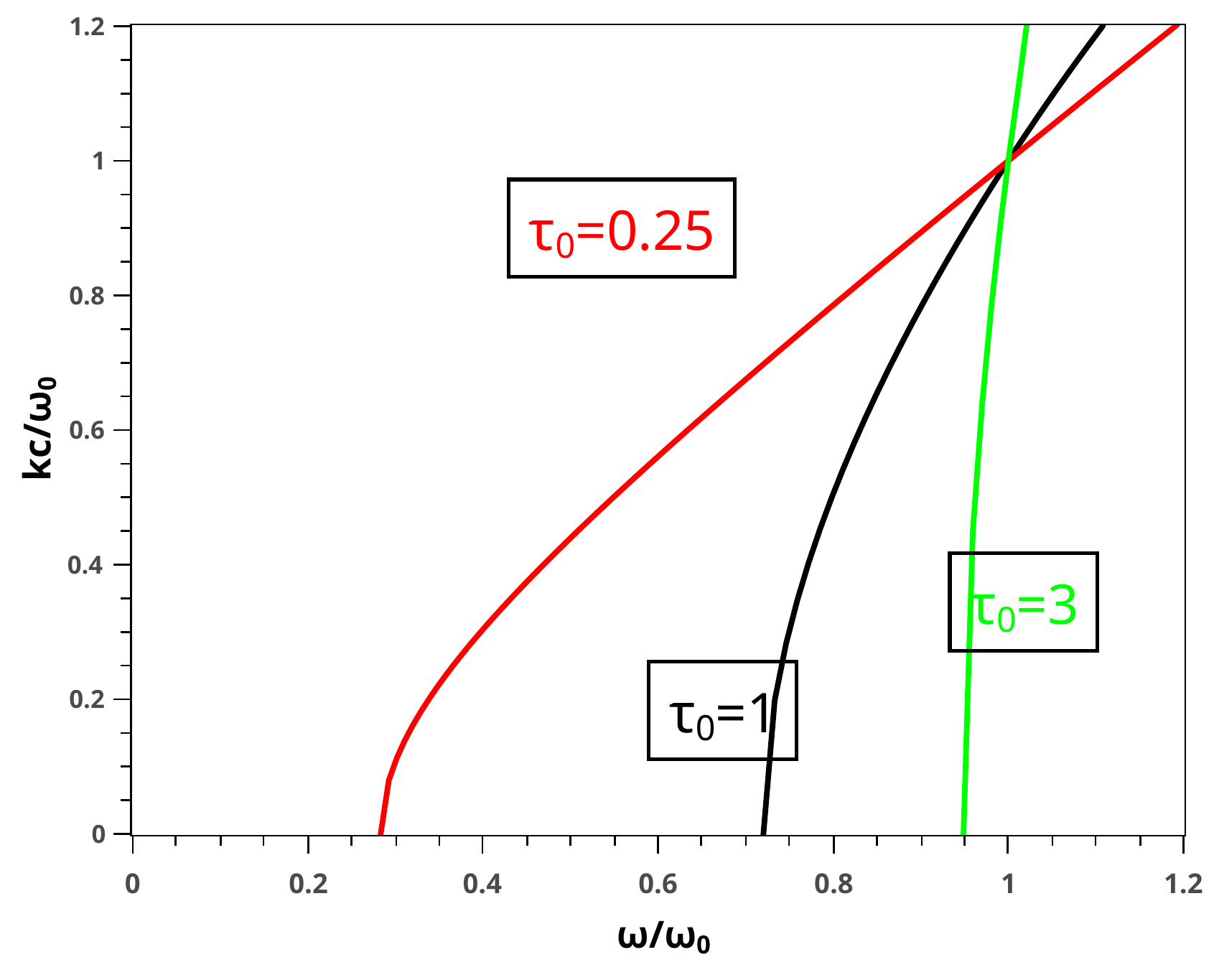}
		\caption{\textit{Sharp line limit}: illustration of the soliton dispersion law $k(\omega)$ for a few different values of scaled pulse width: $\tau_0=\omega_0\tau_p$. Upper pane -- absorbing media $S_0=-1$.Lower pane -- amplifying media media $S_0=1$. }
	\end{center}
\end{figure}
\section{Discussion}
\subsection{Sharp line limit}
In this case line shape function tends to delta function so that: $\tilde\Delta=\Delta$ and $\langle F(\Delta)\rangle=F(\Delta)$ which significantly simplifies further calculations and both, dispersion law and velocity as function of frequency ratio $x=\omega/\omega_0$ attain simple analytic forms:
\barl\label{analyt}
\non K_{\pm}&=&\sqrt{x^2-\frac{2\nu S_0(1-x)\tau^2_0}{1+(1-x)^2\tau^2_0}},\\
V_{\pm}&=&\frac{K}{x-\frac{\nu S_0\tau^2_0}{1+(1-x)^2\tau^2_0}}
\ear

We have graphically presented our results on figures (1) and (2) for absorbing ($S_0=-1$) and amplifying ($S_0=1$) initial conditions. 

In both cases, around resonance $\omega/\omega_0\sim 1$ and for short pulses $\tau_0<1$ we observe similar results as those obtained within the linear
approximation $\omega = ck$. That is, for absorbing media, only
subluminal motion is possible $v<c$. While dispersion law attains 
simple linear functional dependence and velocity gradually decrease as a function of duration time. When pulse duration, for a given $\nu$, exceeds some critical value corresponding a minimum of the function
\be\label{crit}
\tau^2_0={\frac{x^2}{2\nu (1-x)-(1-x)^2x^2}},
\ee
sudden vanishing of $k$ is observed when $\omega/\omega_0 \geq 1$. This indicates immediate pulse stopping. For example, $\tau^{crit}_0\sim 1.94$ for $\nu=1$. Above the critical pulse width two branches in dispersion law appear. First branch lies in the interval $0<x<x_1$, while te second one is within $x_2<x<\infty$. $x_1$ and $x_2$ are solutions of the cubic equation $x^3-x^2-2\nu =0$ where $K(X)=0$.

Such behavior of dispersion law is reflected to the pulse delay, i.e, $v/c$ dependence, as follows: for  $\omega/\omega_0 <1$ we observe expected behavior similar to that in resonance case. That is,  $v<c$ always and slowly decreases as a function of pulse duration. However, for $\omega/\omega_0\equiv x_1 >1$ velocity suddenly vanishes when pulse duration approaches $\tau^{crit}_0$. Nevertheless, for $\omega/\omega_0<1$ velocity still decreases as a function of duration time, but now towards some finite value and can never be stopped. When $\omega/\omega_0 < 0.5 $ pulse velocity becomes practically constant. Out of resonance but for $\omega>\omega_0$ pulse velocity vanishes for large pulses. As indicated by the behavior of $K(x)$ when $x$ exceeds unity pulse \textit{stopping} occurs. 

In the case of amplifying media $S_0=1$ super--luminal motion is predicted  as in the case of resonance. However, in contrast to the resonant case, here there exists a particular lowest value of frequency ratio ($\omega /\omega_0 \sim 1/3 + (\nu/6)\tau^2_0\langle F(\Delta)\rangle$) below which the pulse velocity tends to  sub--luminal region, which contradicts to preceding discussion that for $S_0$ only super--luminal pulses exist.  For intermediate values of frequency ratio $v(\tau_0)> c$ exhibits the expected behavior: increases with $\tau_0$ until it reach critical value, different for each $\omega/ \omega_0$, when the sudden drop is observed. 
\begin{figure}[h]
	\begin{center}
		\includegraphics[height=6 cm]{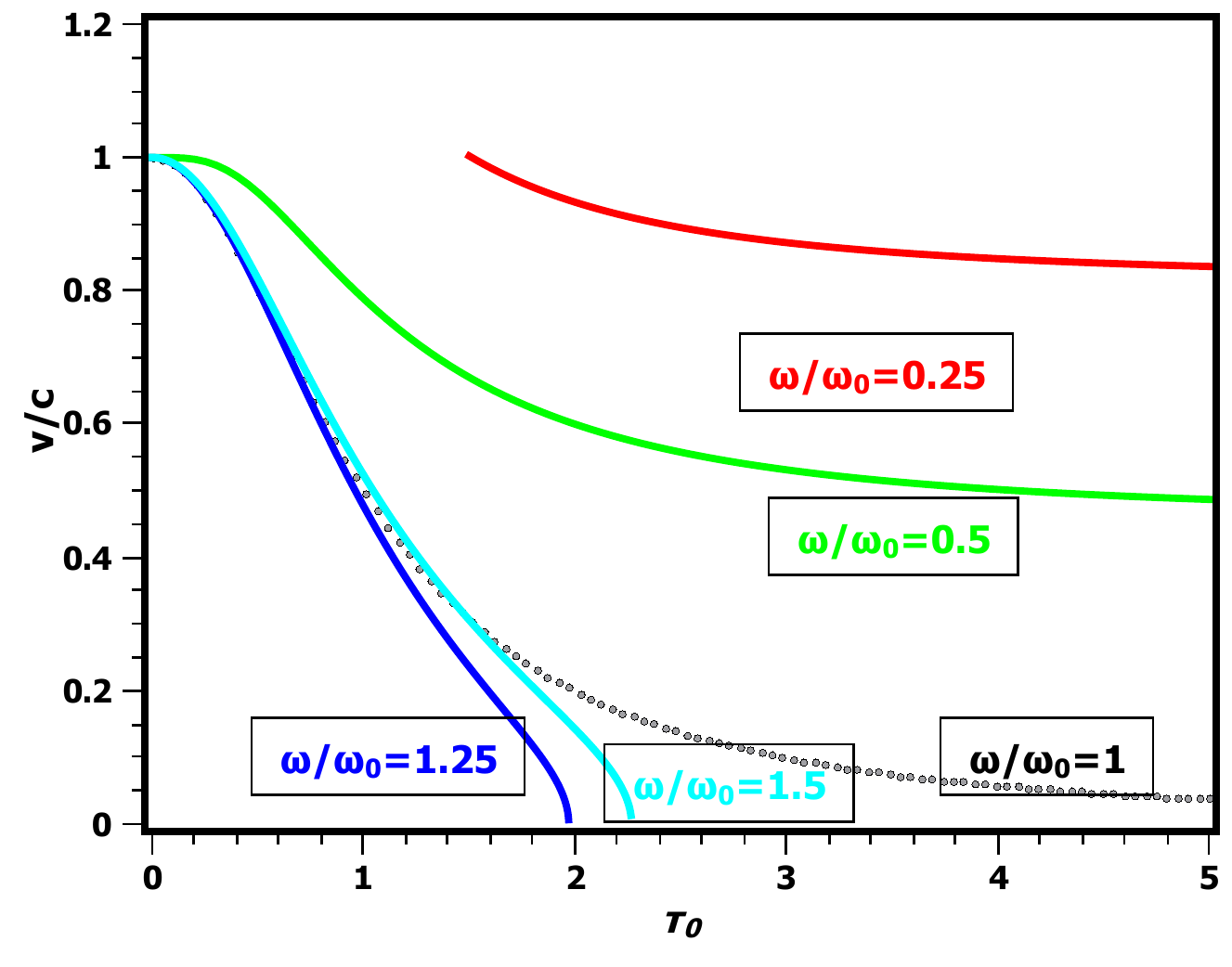}
		\includegraphics[height=6 cm]{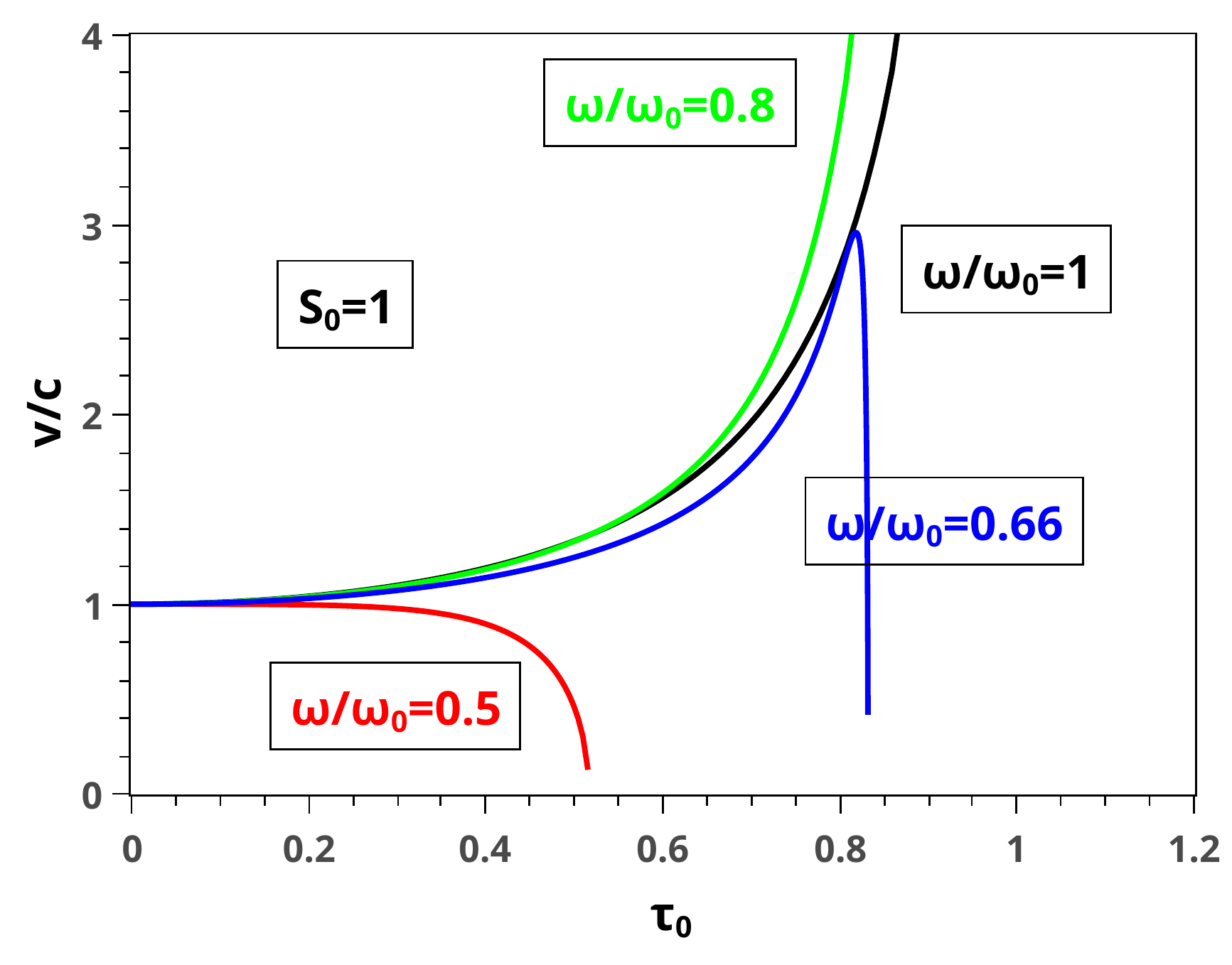}
		\caption{\textit{Sharp line limit}: Pulse velocity delay $v/c$ versus dimensionless pulse width $\omega_0\tau_p$ for a few values of the frequency ratio. Upper pane -- absorbing media $S_0=-1$. Lower pane -- amplifying media $S_0=1$. }
	\end{center}
\end{figure}
\subsection{Influence of the inhomogeneous broadening}
For simplicity we took  line shape function in the Lorentzian form:
\be\label{lsf}
\mathcal{G}(\Delta)=\frac{2\tau^*}{\pi}\frac{1}{1+\Delta^2\tau^{*2}}
\ee
where $\tau*$ stays for the inhomogeneous broadening relaxation time. This choice enables us easy analytic evaluation of the average values in the expressions for dispersion law and velocity delay: 
\barl\label{exact}
\non\langle F(\Delta)\rangle&=&\frac{1}{1+\frac{\tau_p}{\tau*}},\\
\langle \Delta F(\Delta)\rangle&=&\frac{2}{\pi\tau*}\frac{\ln \frac{\tau*}{\tau_p}}{1-\frac{\tau^2_p}{\tau^{*2}}},\\
\non\tilde\Delta&=&\frac{2}{\pi\tau^*}\frac{\ln \frac{\tau^*}{\tau_p}}{1-\frac{\tau_p}{\tau^*}}.
\ear
Accordingly, dispersion law and pulse delay became:
\barl\label{ihb}
\non K_{\pm}&=&\sqrt{x^2+\frac{4S_0\nu \omega_0\tau*}{\pi}y^4\frac{\ln y}{y^2-1}}, \;\; y=\frac{\tau_p}{\tau^*},\\
V_{\pm}&=&\frac{K_{\pm}}{x-\frac{\nu S_0(\omega_0\tau*)^{2}y^2}{1+y}}
\ear
Our results  are visualized in Fig.(3). In upper pane we have plotted, in dimensionless units, dispersion law as function of frequency ratio ($x=\omega/\omega_0$)  for a few different values of ratio of the pulse width over the inhomogeneous broadening relaxation time -- $y=\tau_p/\tau^*$. In lower pane we have presented pulse velocity as function of  $y=\tau_p/\tau^*$. Dispersion law exhibits substantially different behavior with respect to that observed within the sharp line limit. This particularly concerns the near resonance case ($\omega/\omega_0 \sim 1$) where our results do not tend to the known ones obtained in the resonance. This is the consequence of the appearance of the constant shifts, determined by the ratio $y$, in the expressions for dispersion law and pulse delay (\ref{ihb}). In both cases these shifts tend to zero when $\tau^*\gg \tau_p$ and our results approach those obtained within the strict resonance. 

Dispersion relation for finite values of $y$  exhibits very specific behavior for absorbing and amplifying  media. That is, for absorbing media, for each particular value of $y$ there is a minimal value of frequency ratio ($x=\omega/\omega_0$)  below which there is no solutions for $K$, that is pulse does not exist. When   $x$ exceeds this minimal value $K(x)$ monotonically increases approaching linear dependence for large $x$. For large $y$ this limit is approached for the un--physically large values of $x\gg 1$ when presented theory of SIT does not hold.

For the amplifying media, starting from some minimal value, $K$ monotonically increases approaching, again, linear dependence for large $x$. In contrast to absorbing media pulse   exists for all $x$.

In the absorbing media, for each $x$, pulse velocity exhibits similar behavior as well as in the sharp line limit: as a function of $y$  it gradually decay towards zero approaching it for some critical value specific for each $x$. 

In the amplifying media super--luminal pulse motion is predicted: for each the particular value of the frequency ratio, $v(y)$ exhibits qualitatively the same behavior as well as in the case of resonance with no accounted for effects of dispersion (represented with curve constructed of  $\diamond$). That is, as a function of $y$, $v/c$ monotonically increases from unity to infinity, while each curve may be recovered from the some particular one, say $x=1$, by simply rescaling frequency ratio.  
\begin{figure}[h]
	\begin{center}
		\includegraphics[height=6 cm]{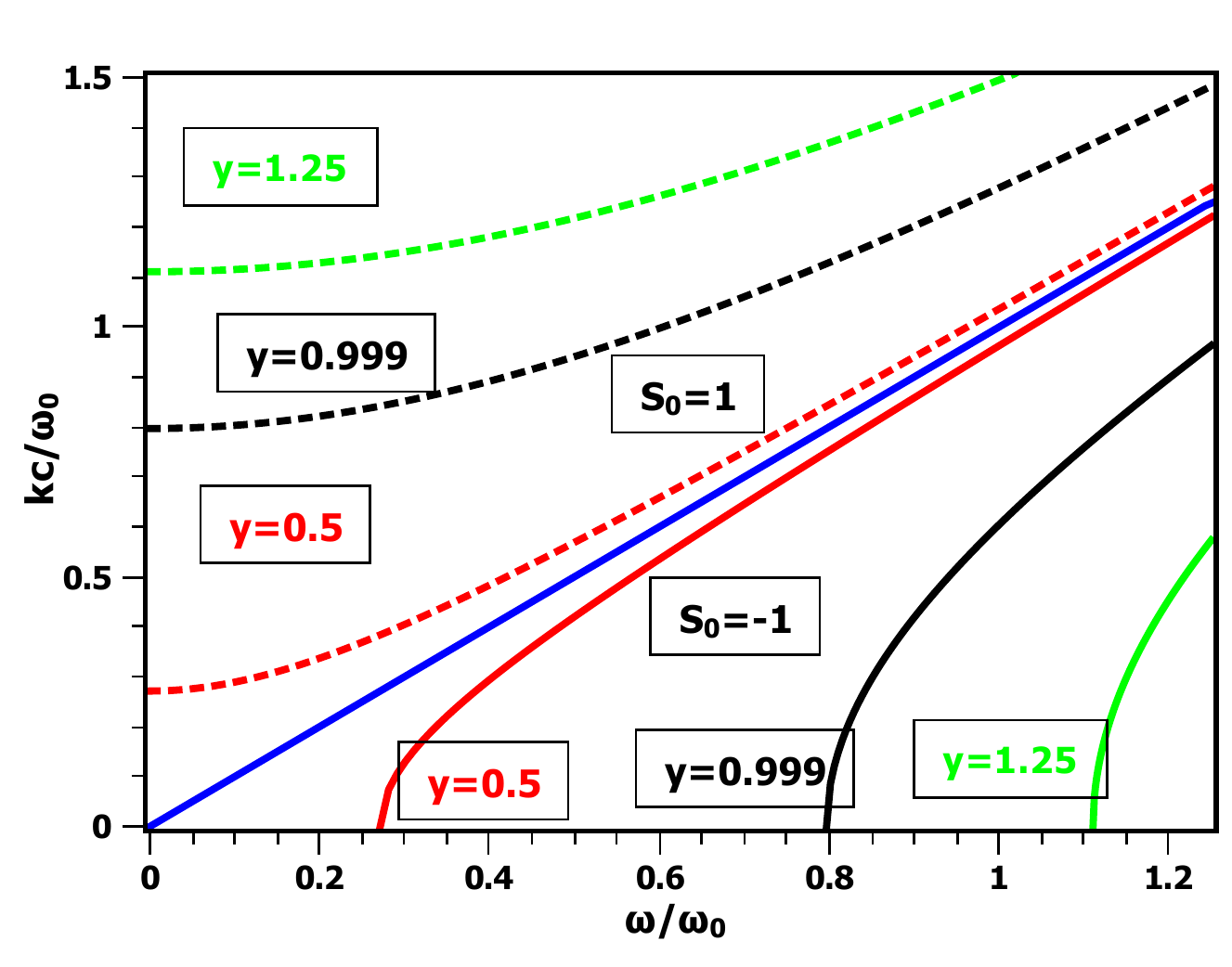}
		\includegraphics[height=6 cm]{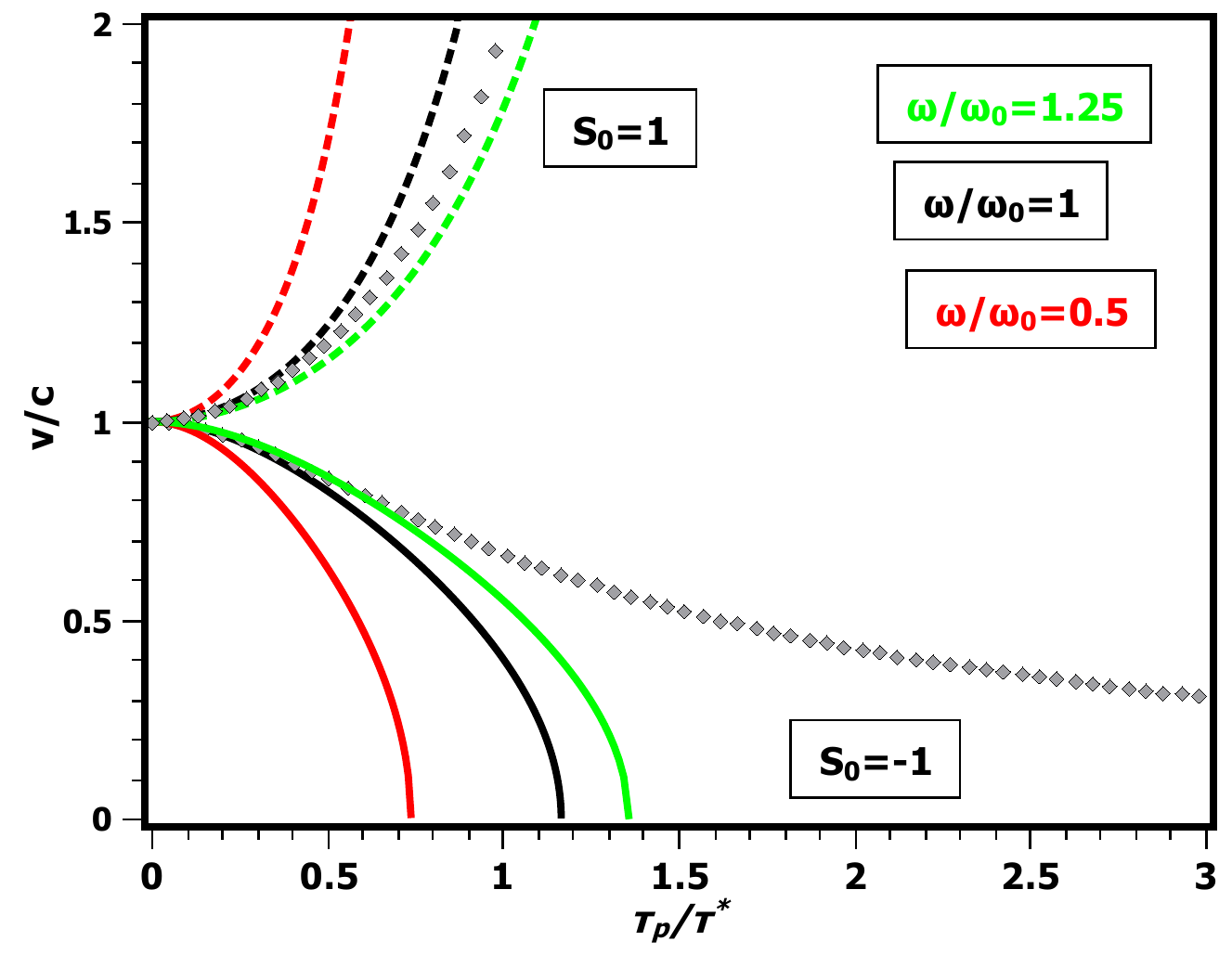}
		\caption{\textit{Illustration of the impact of the inhomogeneous broadening on SIT pulse properties}: Upper pane -- carrier wave quasi--momentum versus frequency ratio. Dispersion relation for absorbing media ($S_0=-1$) is visualized by full lines, while to amplifying media correspond dotted lines. Blu full line corresponds to resonance case $\omega = kc$. Lower pane -- pulse velocity delay. Full lines correspond to absorbing media $S_0=-1$. Dotted lines correspond to amplifying media. Curves indicated by $\diamond$ stand for resonance.}
	\end{center}
\end{figure}
\section{Concluding remarks}
Our study reveals some new features of the SIT phenomenon stemming from the spatial dispersion of the carrier wave $\omega(k)$. We found  that the properties of the SIT pulse substantially depends on the frequency ratio ($x=\omega/\omega_0$). This particularly concerns the delay of the pulse velocity which may be controlled by means of varying of $x$. In that sense the most interesting consequence is the possibility of full stopping of SIT pulse. In particular, in the sharp line limit, for each value of $x>1$ EM pulse of is fully stopped  (absorbed) by  the medium provided that its width (duration time) exceeds some critical value. 
In the case of inhomogeneously broadened media each EM pulse gets stopped irrespectively on the value of $x$, that is there is no any limitation on the value of $x$ which may be arbitrarily low provided that pulse is wide enough.  

In order to relate dispersion with absorption coefficient we derive the Area theorem from the system 
(\ref{SE02B}) 
\be\label{at}
\frac{\partial \theta}{\partial x}=\frac{\beta }{2} \sin \theta,\;\; \beta=\frac{\gamma\omega^2_0\mathcal{G}(0)}{\kappa^2c^2 k}\equiv \frac{2\gamma\omega^2_0\tau^*}{\kappa^2c^2 k}.
\ee
Apparently, absorption coefficient due to $1/k$ dependence tends to infinity, indicating its full absorption for $k\mapsto 0.$ 

In conclusion, our results point to possible new means of the control of propagation of EM waves. It relies on the prediction of possible dramatic influence of the frequency ratio $x=\omega/\omega_0$ on carrier wave of SIT pulse and its velocity whose vanishing may be expected for a convenient choice of $x$ and pulse duration time. In systems built of natural atoms, this may not be easily realized due to small values of material constant. Nevertheless, tunability of the parameters of artificial atoms may enhance the predicted effect and make it possible.  Also, by applying an additional driving field like in EIT, mixed induced transparency (SIT+EIT), may be achieved where the best features of both effects were exploited as shown in \cite{park}.

\begin{acknowledgments}
This work was partially supported by the Ministry of Education, Science and Technological Development of Republic Serbia, Grants No. III - 45010 and OI - 171009, the Ministry of Science and Higher Education of the Russian Federation in the framework of Increase Competitiveness Program of NUST "MISiS" (No. K2-2019-010), implemented by a governmental decree dated 16th of March 2013, N 211. NL also acknowledges support by General Secretariat for Research and Technology (GSRT) and the Hellenic Foundation for Research and Innovation (HFRI) (Grant no.: 203).
\end{acknowledgments}

\end{document}